\documentclass[aps,nofootinbib,groupedaddress]{revtex4}
\usepackage{graphicx}
\usepackage{amsmath,amssymb}
\usepackage{hyperref}
\usepackage[usenames]{color}

\hypersetup{
    colorlinks=true,
    linkcolor=red,
    citecolor=blue,
}

\def\be{\begin{equation}}
\def\ee{\end{equation}}
\def\ba{\begin{eqnarray}}
\def\ea{\end{eqnarray}}

\newcommand\ie{{\em i.e. }}

\begin{document}

\title{Primordial Magnetism in the CMB: Exact Treatment of Faraday Rotation 
and WMAP7 Bounds}

\author{
Levon Pogosian$^{1}$, 
Amit P.S. Yadav$^{2}$,
Yi-Fung Ng$^{3}$ 
and Tanmay Vachaspati$^{4}$
}

\affiliation{
$^1$Department of Physics, Simon Fraser University, Burnaby, BC, V5A 1S6, Canada \\
$^2$Institute for Advanced Study, Princeton, NJ 08540, USA\\
$^3$CERCA, Physics Department, Case Western Reserve University, Cleveland, OH 44106-7079, USA\\
$^4$Physics Department, Arizona State University, Tempe, AZ 85287, USA  \\
}

\begin{abstract}
Faraday rotation induced B-modes can provide a distinctive signature 
of primordial magnetic fields because of their characteristic frequency 
dependence and because they are only weakly damped on small scales, 
allowing them to dominate B-modes from other sources. By numerically 
solving the full CMB radiative transport equations, we study the B-mode 
power spectrum induced by stochastic magnetic fields that have significant 
power on scales smaller than the thickness of the last scattering surface. 
Constraints on the magnetic field energy density and inertial scale are 
derived from WMAP 7-year data, and are stronger than the big bang 
nucleosynthesis (BBN) bound for a range of parameters. Observations 
of the CMB polarization at smaller angular scales are crucial to provide 
tighter constraints or a detection.
\end{abstract}

\maketitle

\section{Introduction}
\label{sec:introduction}

Many early universe scenarios predict the existence of cosmological magnetic 
fields and several observational techniques are currently being employed to 
detect such fields~\cite{RevModPhys.74.775,Kandus:2010nw}. There are also recent claims
for the detection of an inter-galactic magnetic 
field~\cite{Neronov:1900zz}.
With further confirmation and refinement, these observations can be turned 
into a probe of the fundamental interactions that are necessary to generate 
a primordial magnetic field and to the physics of the early universe. 
A primordial magnetic field can also have important implications for 
the formation of first stars, growth of cosmic structure and the present 
universe. 

Of particular interest to us are magnetogenesis scenarios based on phase 
transitions such as the electroweak phase transition, when Maxwellian 
electromagnetism first emerged  
\cite{Vachaspati:1991nm,Cornwall:1997ms,Vachaspati:2001nb,
GarciaBellido:2003wd,Copi:2008he,Vachaspati:2008pi,Ng:2010mt,Chu:2011}. 
In this scenario  particle physics uncertainties are minimal and 
magnetogenesis is tightly 
related to the creation of matter, or ``baryogenesis'' -- the magnitude
of the magnetic helicity density is approximately equal to the baryon number
density. Since the cosmic number density of baryons is known, the scenario 
enables a prediction for the magnetic helicity density that is largely 
independent of the details of the electroweak model. Furthermore, the
{\it left-handed} magnetic helicity is a direct outcome of parity violation
present in the electroweak model and, like baryon number, is a remarkable 
large-scale manifestation of a microscopic symmetry violation.

Non-vanishing helicity has important consequences for the evolution of a
magnetic field. Although the field is generated on small scales, magnetic
helicity allows for an ``inverse cascade'' where power is transferred
from small to large scales, resulting in magnetic coherence on larger
scales. In the case of magnetic fields generated at the electroweak
scale, the final coherence scale can be on the parsec to kiloparsec
scales \cite{Vachaspati:2001nb,Sigl:2002kt}. While the coherence scale 
is large, it is not as large as
for fields that might be generated during an inflationary epoch.
For purposes of calculating observational signatures, magnetic fields
generated at a phase transition are stochastic. An important aim of
the present work is to find distinctive observational signatures of
magnetic fields that are stochastic on sub-Mpc coherence scales.

\begin{figure}
  \includegraphics[height=0.45\textwidth,angle=-90]{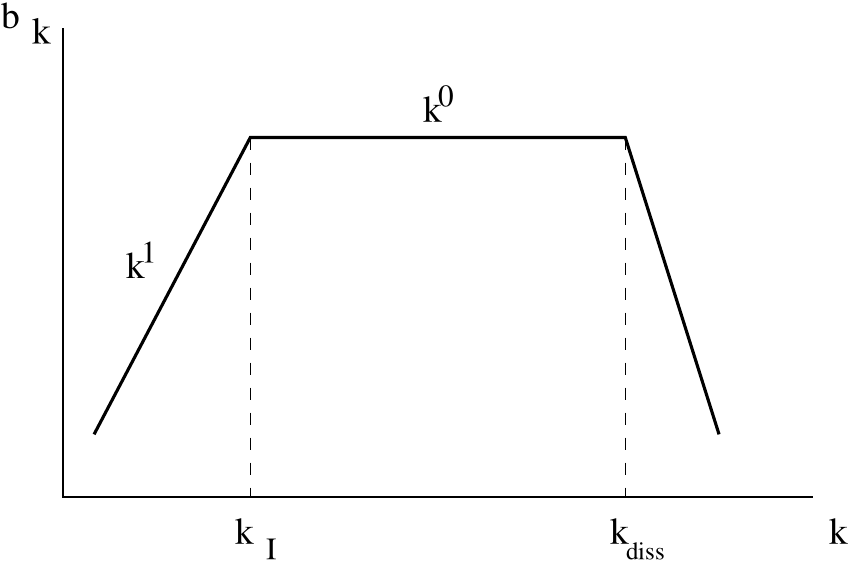}
  \caption{The Fourier amplitude $b(k)$ -- as given by the square root of the
power spectrum -- of a stochastic magnetic field generated
during a cosmic phase transition is expected to grow for 
$k < k_I$ where $k_I$ is an inertial scale, and obey some other
power law for $k_I < k < k_{\rm diss}$, where $k_{\rm diss}$ is a 
dissipative scale. The work of \cite{Jedamzik:2010cy} suggests
$k^1$ growth at small $k$, then $k^0$ behavior until $k_{\rm diss}$,
beyond which the amplitude falls off very quickly.}
  \label{Bspectrum}
\end{figure}

Once magnetic fields have been injected into the cosmological plasma, 
the subsequent evolution is described by magneto-hydrodynamical (MHD) 
equations in an expanding spacetime. Power 
on very small length scales is expected to be erased by dissipational 
mechanisms. Power on very large scales is also suppressed because 
the magnetic field is injected on microscopic scales and cannot extend 
to arbitrarily large scales. Hence the spectral distribution of the 
magnetic field is expected to decay fast on small and large scales and
be peaked on some intermediate scale, which is presumably at the parsec
to kiloparsec scale. The spectral form of the magnetic
field has been investigated recently in some detail in 
Refs.~\cite{Jedamzik:2010cy,2004PhRvD..70l3003B} and the results 
are schematically depicted in Fig.~\ref{Bspectrum}.

On the observational front, magnetic fields within galaxies and clusters 
of galaxies have been studied for many decades and their origin --
cosmological versus astrophysical -- remains unsettled, though a hybrid
explanation is also conceivable. More recently, observations of TeV gamma 
ray sources have been used to place {\it lower} bounds $\sim 10^{-16}$ G 
\cite{Neronov:1900zz,Tavecchio:2010ja,Dolag:2010ni,Taylor:2011bn,
Takahashi:2011ac}, and perhaps a measurement $\sim 10^{-15}$ G 
\cite{Ando:2010rb}, on a magnetic field in inter-galactic space, $\sim 10$ Mpc 
away from the TeV gamma ray source. It seems likely that such an 
inter-galactic magnetic field, unassociated with cosmic structure, 
is primordial, but an astrophysical origin, say based on the expulsion 
of magnetic fields from active sources, may also be viable.

A detection of magnetic fields in the cosmic microwave background (CMB), 
for example due to Faraday rotation (FR) of the CMB polarization, would 
unambiguously 
point to a cosmological origin because there are no confounding magnetized 
structures at last scattering. However, an observed FR
of the CMB could also be due to magnetic fields along the line of
sight, especially within the Milky Way. Hence, it is necessary to find 
distinctive signatures of FR that
occurred at recombination versus that which happened more recently. 
As we will see, in addition to its characteristic frequency dependence, 
FR induced B-modes are only weakly damped on small angular scales 
(high $\ell$), which means they are likely to dominate B-modes from 
other sources. 

Earlier work on FR of the CMB in cosmic magnetic fields has largely 
focused on the effect of a uniform magnetic field 
\cite{Kosowsky:1996yc,Harari:1996ac} and, when a stochastic magnetic 
field has been considered, a ``thin'' LSS was often assumed 
\cite{Kosowsky:2004zh,Giovannini:2008aa,Kahniashvili:2008hx}. This approximation results 
in a tremendous technical simplification but it is not clear if 
it is suited to study the effects of a magnetic field with coherence 
scale smaller than the thickness of the LSS. In \cite{Giovannini:2008jv}, the finite thickness effects of the LSS were modelled by approximating 
the visibility function with a Gaussian profile. As we show, the thin LSS approximation is 
sufficient for order of magnitude estimates, but can be wrong by factors 
of a few and in an $\ell$ dependent way. 

In the present paper, we study the effect of a primordial magnetic field
on the CMB polarization, focussing on the effect of a stochastic field
with power on small (sub-Mpc) scales. We calculate the B-mode
correlator, $C_l^{BB}$, induced by a primordial magnetic field. In the thin
LSS approximation of Sec.~\ref{sec:thin}, 
we show that $C_l^{BB}$ is directly related to $C_l^{EE}$ multiplied 
by the correlation function for the FR, 
$C_l^{\alpha\alpha}$, which we calculate in 
Sec.~\ref{sec:magnetic-spectrum}. In Sec.~\ref{sec:exactthick} we move 
on to our main calculation of $C_l^{BB}$ with a thick LSS.
Here we find that the physics of FR during recombination 
can be described in terms of ``window functions'' through which the 
magnetic field spectrum (see Eq.~(\ref{ClBBthick})) appears in 
$C_l^{BB}$. We have to resort to extensive numerical efforts to compute 
the window functions. Our results are described in Sec.~\ref{results}. 
The window functions themselves are independent of the magnetic field 
power spectrum and are shown in Fig.~\ref{fig:windows}. When the
window functions are convolved with the magnetic field power spectrum, 
we obtain $C_l^{BB}$. In Figs.~\ref{fig:cl0}, \ref{fig:cl5} we show 
the results for $C_l^{BB}$ for a scale-invariant and a blue magnetic 
power spectrum. We conclude in Sec.~\ref{conclusions}.  We have made our window functions publicly 
available at {\tt http://www.sfu.ca/$\sim$levon/faraday.html} along with a short Fortran code for calculating 
$C_l^{BB}$.

\section{Stochastic Magnetic Fields}
\label{sec:B}

A statistically homogeneous and isotropic stochastic magnetic field 
is described by the two-point correlator in Fourier space 
as 
\be
\langle b_i ({\bf k} ) b_j ({\bf k}' ) \rangle =
(2\pi )^3 \delta^{(3)}({\bf k} + {\bf k}' )
[ (\delta_{ij} - {\hat k}_i {\hat k}_j) S(k) +
i \varepsilon_{ijl} {\hat k}_l A(k) ] \ ,
\label{bcorr}
\ee
where $S(k)$ and $A(k)$, the symmetric and anti-symmetric magnetic
power spectra, are real functions of $k=|{\bf k}|$. Throughout the paper, we use Gaussian CGS units.
The energy density in modes up to some value of $k$ is given by
\begin{equation}
\epsilon_M (k) = \frac{1}{8\pi} \int_ {k'<k} \frac{d^3k'}{(2\pi)^3} 2S(k') 
               = \frac{1}{(2\pi)^3} \int_0^{k} dk k^2 S(k) . 
\label{epsBk}
\end{equation}

We take the form of $S(k)$ to correspond to Fig.~\ref{Bspectrum} 
\begin{eqnarray}
S(k) = \begin{cases}
S_* \left ( \frac{k}{k_I} \right ) ^{2n-3}, &\mbox{$0<k<k_I$}\\
S_* \left ( \frac{k}{k_I} \right ) ^{2n'-3}, \ 
                  &\mbox{$k_I< k< k_{\rm diss}$}\\
0, & \mbox{$k_{\rm diss} < k$}
       \end{cases}
\end{eqnarray}
The results in \cite{Durrer:2003ja,Jedamzik:2010cy} suggest the exponents $n=5/2$ 
and $n'=3/2$. In our analysis, along with these values, we will also consider the case of
a nearly scale invariant spectrum, with $2n=2n'=0.1$, motivated by an inflationary mechanism 
of generation of magnetic fields \cite{Turner:1987bw,Ratra:1991bn}.

We will only consider the effect of magnetic field on the CMB for 
modes with $l \le l_{\rm max} = 10^4$, as computations at higher $l$ 
are very expensive. This corresponds to 
a minimum comoving scale of 1~Mpc, or 
$k_{\rm max} \sim 1 ~{\rm Mpc}^{-1}$. For magnetic fields generated 
at the electroweak phase transition, the coherence scale is 
estimated at kpc scales or less \cite{Vachaspati:2001nb,Sigl:2002kt}. 
Hence $k_I$ may be expected to be $10^{3} ~ {\rm Mpc}^{-1}$. This is
much larger than $k_{\rm max}$ and thus the $l \le 10^4$ modes 
of the CMB are likely to be affected only by the magnetic field modes in the 
inertial range $0 < k < k_I$, and the form of the power spectrum
for $k > k_I$ plays no direct role. However, even then, the large $k$ part of 
the power spectrum would still enter when we derive a constraint on the
magnetic field, since the amplitude, $S_*$, is fixed by the total 
energy density in the magnetic field. We now make this point clearer.

First define an ``effective magnetic field'', $B_{\rm eff}$, in
terms of the  total energy density in the magnetic field, $\epsilon_0$,
\begin{equation}
\epsilon_0 \equiv \frac{B_{\rm eff}^2}{8\pi} \ .
\label{def-beff}
\end{equation}
In other words, $B_{\rm eff}$ is the field strength of a uniform
magnetic field that has the same total energy density as our
stochastic magnetic field. Our constraints will be written in
terms of $B_{\rm eff}$. To connect to the amplitude of the
power spectrum, we first evaluate the energy density in the
magnetic field. From Eq.~(\ref{epsBk}) we get
\begin{eqnarray}
\epsilon_M (k) = \begin{cases}
      \frac{S_* k_I^3}{16\pi^3 n} 
                   \left ( \frac{k}{k_I} \right )^{2n} ,
                       &\mbox{$k \le k_I$} \\
                \frac{S_* k_I^3}{16\pi^3 n} \left [ 1 + 
  \frac{n}{n'} \left \{ \left ( \frac{k}{k_I}  \right )^{2n'} 
           -1 \right \} \right ] , 
                 &\mbox{$k_I < k \le k_{\rm diss}$} \\
                \frac{S_* k_I^3}{16\pi^3 n} \left [ 1 + 
  \frac{n}{n'} \left \{ \left ( \frac{k_{\rm diss}}{k_I}  \right )^{2n'} 
           -1 \right \} \right ] , 
                 &\mbox{$k_{\rm diss} < k$}
                  \end{cases}
\end{eqnarray}
The total energy density in the magnetic field, $\epsilon_0$,
is found by setting $k \to \infty$, which is the same as $\epsilon_M (k)$ for $k>k_{\rm diss}$. Thus, we can write
\be
\epsilon_0 = \frac{S_* k_I^3 \kappa}{16\pi^3 n} \ ,
\ee
where
\be
\kappa \equiv 1 + 
  \frac{n}{n'} \left \{ \left ( \frac{k_{\rm diss}}{k_I}  \right )^{2n'}
           -1 \right \}  \ ,
\label{kappadef}
\ee
and hence
\begin{equation}
B_{\rm eff} = \frac{1}{\pi}\sqrt{\frac{\kappa S_* k_I^3}{2n}} \ .
\end{equation}
For a fixed exponent $n$, CMB observations at $l \le 10^4$ will only 
constrain the combination $S_* k_I^3$. To then convert the constraint 
to a bound on the energy density in magnetic fields requires knowledge 
of the exponent $n'$ and the inertial and dissipation scales. 
In other words, the CMB signature for $l \le 10^4$ probes the long
wavelength tail of the magnetic spectrum and not the modes where the 
bulk of the energy density resides. This suggests that it may be 
favorable to investigate the CMB at yet higher $l$; indeed, our 
results do show stronger signatures with growing $l$. 

Big bang nucleosynthesis (BBN) is sensitive to the total
energy density in the magnetic field since this is what enters
the expansion rate of the universe. The constraint from BBN 
~\cite{Matese_OConnell_69,1996PhRvD..54.7207K,Grasso:1996kk,
Cheng:1996yi} 
is best expressed in terms of the magnetic field energy density relative
to the photon energy density
\begin{equation}
\Omega_{B\gamma} \equiv \frac{\epsilon_0}{\rho_\gamma} \lesssim 10^{-1}
\end{equation}
The relative density, $\Omega_{B\gamma}$, is independent of cosmological
epoch since both magnetic field and photon energy density scale
as $a^{-4}$ where $a(t)$ is the cosmological scale factor. 
Note that $\rho_\gamma$ is the photon density which is different from
the radiation density at BBN epoch since neutrinos also contribute to
radiation.

Finally, all the scaling due to the expansion of the universe
can be pulled out by converting to rescaled quantities
\begin{equation}
{\bf B} a^2 \rightarrow {\bf B}  \ , \ \ 
\rho_\gamma a^4  \rightarrow \rho_\gamma \ , \ \ 
k a \rightarrow k  \ , 
\end{equation}
and, in what follows, we will use these comoving quantities unless 
explicitly stated. The final form of the magnetic field power spectrum 
in the inertial range is
\begin{equation}
S(k)  = 
\Omega_{B\gamma}  \rho_\gamma 
              \frac{16\pi^3 n}{\kappa k_I^3}
              \left ( \frac{k}{k_I} \right )^{2n-3} \ , \ \ 
{k} < {k_I}
\label{tildeSk}
\end{equation}

The power spectrum $S(k)$ will enter the calculation of $C_l^{BB}$ in
combination with powers of $k$. So it is convenient to introduce
the dimensionless ``power spectrum'' using powers of $k$ and also
the wavelength of observed radiation, $\lambda_0$,
\begin{equation}
\Delta^2_M(k) \equiv k^3 S(k) \left( \frac{3 \lambda_0^2}{16 \pi^2 e} \right)^2 
 = \begin{cases}
 \Delta^2_0  \left ( \frac{k}{k_I} \right )^{2n} &\mbox{$0<k<k_I$}\\ 
 \Delta^2_0  \left ( \frac{k}{k_I} \right )^{2n'} &\mbox{$k_I<k<k_{\rm diss}$}\\
 0 &\mbox{$k>k_{\rm diss}$}
 \end{cases}
\label{DeltaB2}
\end{equation}
where
\begin{equation}
\Delta^2_0 \equiv \frac{9n}{16\pi e^2 \kappa} \rho_\gamma {\lambda_0^4}  
              ~  \Omega_{B\gamma} 
\end{equation}
At the present epoch 
$\rho_\gamma (t_0) = 4.64 \times 10^{-34} {\rm gm/cm^3} = 2 \times 10^{-15} {\rm (eV)}^4$ 
\cite{dodelson}, and so
\begin{equation}
\Delta^2_0 = 1.1\times 10^4 ~ \frac{\Omega_{B\gamma}}{\kappa} \times 
  \left ( \frac{2n}{5} \right ) \left ( \frac{90~{\rm GHz}}{\nu_0} \right )^4
\label{Delta20}
\end{equation}
where we denote the observed CMB frequency by $\nu_0$. Note that $\Delta^2_0$ 
is independent of $k_I$.

In principle, $k_{\rm diss}$ is not an independent parameter. One can estimate 
its value for a given amplitude and shape of the magnetic fields 
spectrum. According to \cite{Jedamzik:1996wp,Kahniashvili:2010wm}, 
$k_{\rm diss}$ is determined by damping into Alfven waves and can be 
related to $B_{\rm eff}$ as
\be
%{k_{\rm diss} \over 1{\rm Mpc}^{-1}} = 
%          1.4 ~ \sqrt{(2\pi)^{2n}h \over \Gamma(n+1)} 
%          \left( 10^{-7} {\rm Gauss} \over B_{\rm eff} \right) \ ,
{k_{\rm diss} \over 1{\rm Mpc}^{-1}} \approx
          1.4 \ h^{1/2} \left( 10^{-7} {\rm Gauss} \over B_{\rm eff} \right) \ ,
\label{kIBeff}
\ee
Converting this to $\Omega_{B\gamma}$, we obtain
\be
%k_{\rm diss} \approx 0.43 \sqrt{ {(2\pi)^{2n}h \over \Gamma(n+1)} 
%\left( {10^{-2} \over \Omega_{B\gamma}} \right)}
%\ {\rm Mpc}^{-1} \ ,
k_{\rm diss} \approx 0.43 \sqrt{10^{-2} h \over \Omega_{B\gamma}}
\ {\rm Mpc}^{-1} \ ,
\label{kIOmegaB}
\ee
where it was useful to know that $1~{\rm Gauss} = 6.9\times 10^{-20}~{\rm GeV}^2$ 
when converting between natural units and CGS units. We also note that Eq.~(\ref{def-beff}) implies:
\be
B_{\rm eff}= 3.25 \times 10^{-6} \sqrt{\Omega_{B \gamma}} \ {\rm Gauss} \ .
\label{beff-omega}
\ee

The relation (\ref{kIBeff}) is based on the analysis in 
Ref.~\cite{Jedamzik:1996wp} where small perturbations on top of a homogeneous
magnetic field were treated. To extend this analysis to a stochastic magnetic 
field with little power on long wavelengths, Ref.~\cite{Kahniashvili:2010wm} 
introduced a smoothing procedure and split the spectrum into a ``homogeneous'' 
part and a ``perturbations'' part. 
It is not clear to us if this procedure is valid for an arbitrary spectrum,
$S(k)$, but we will still use Eq.~(\ref{kIBeff}) as an approximate expression 
for the dissipation scale.

\section{Faraday Rotation Correlators}
\label{sec:magnetic-spectrum}

The CMB is linearly polarized and an intervening magnetic field 
will rotate the polarization vector at a rate given by:
\begin{equation}
d\alpha = 
\lambda^2 \frac{e^3}{2\pi m_e^2} a \ n_e {\bf B}\cdot d{\bf l} \ ,
\label{theta1}
\end{equation}
where $\lambda$ is the wavelength of light, 
$a$ is the scale factor normalized so that 
$a_{\text{today}}=1$, $n_e$ is the number density of free electrons, 
$d {\bf l}$ is the comoving
length element along the photon trajectory from the 
source to the observer and we
are using Gaussian natural units with $\hbar=c=1$.
Using the known expression for Thomson scattering cross-section,
\begin{equation}
\sigma_T = \frac{8 \pi e^4 }{3 m_e^2} \ ,
\end{equation}
and integrating along the line of sight, we obtain the Faraday rotation
of the polarization angle, 
\begin{equation}
\alpha = \frac{3}{{16 \pi^2 e}} \lambda_0^2 
\int \dot{\tau}({\bf x}) \ {\bf {\tilde B}} \cdot d {\bf l}
\label{theta2}
\end{equation}
where $\dot{\tau}({\bf x}) \equiv n_e \sigma_T a$ is the differential
optical depth along the line of sight, $\lambda_0$ is the
observed wavelength of the radiation and
${\bf {\tilde B}} \equiv {\bf B} a^2$ is the ``comoving'' magnetic field.
The limits of the integral are from the initial position of the photon
to the final position.

FR depends on the free electron density, which becomes 
negligible towards the end of recombination. Therefore, the bulk of 
the rotation is produced during a relatively brief period of time when 
the electron density is sufficiently low for polarization to be produced 
and yet sufficiently high for the FR to occur.
The average FR (in radians) between Thomson scatterings
due to a tangled magnetic field was calculated in Ref.~\cite{Harari} 
and is given by
\begin{equation}
F = \frac{3}{8 \pi^2 e} \frac{B_0}{\nu_0^2}
\approx 0.08 \left( \frac{B_0}{10^{-9}{\rm G}} \right)
\left( \frac{30{\rm GHz} }{\nu_0} \right)^2 \ ,
\end{equation}
where $B_0$ is the current amplitude of the field and $\nu_0$ is
the radiation frequency observed today.

In this section we will calculate the two-point correlation functions of 
the FR angle, $\alpha$, and this will be related to the 
two-point correlation function of the magnetic field as given in
Eq.~(\ref{bcorr}). The FR correlator will enter the calculation 
of $C_l^{BB}$ in the thin LSS approximation of Sec.~\ref{sec:thin}, 
in which one assumes that all of the polarization was generated at once 
independently from the FR. In the general case, which we present in 
Sec.~\ref{sec:exactthick}, the generation of the CMB polarization 
and its FR are entangled.

FR is sensitive only to the line of sight component of 
the magnetic field, whereas magnetic helicity, 
described by the helical power spectrum $A(k)$ in Eq.~(\ref{bcorr}), 
depends on all 3 components of the magnetic field. So a correlator 
of FR cannot be sensitive to the helical properties \cite{Ensslin:2003ez,Campanelli:2004pm,Kosowsky:2004zh}
\footnote{Indirectly though, helicity does
enter the FR signature because magnetic helicity
plays a crucial role in the evolution of magnetic fields and 
the exponent $n$ in $S(k)$ (see Eq.~(\ref{tildeSk})).}.
Following Ref.~\cite{2002PhRvD..65h3502P} without the helical term, we get
\begin{eqnarray}
\langle \alpha ({\hat n}) \alpha ({\hat n}') \rangle &=& 
\left(\frac{3\lambda_0^2}{16 \pi^2 e} \right)^2
\int \frac{d^3k}{(2\pi )^3} S(k) \int d\eta \int d\eta'
 {\dot \tau}(\eta) {\dot \tau}(\eta') 
    e^{-i{\bf k}\cdot {\hat n}\eta} e^{+i{\bf k}\cdot {\hat n}'\eta'}
[ {\hat n}\cdot {\hat n}' - ({\hat k}\cdot {\hat n})( {\hat k}\cdot {\hat n}')] 
\label{alpha_n_npr}
\end{eqnarray}
where ${\hat n},~{\hat n}'$ are two directions on the sky. Note that,
as is conventional, we have written $\dot \tau ({\bf x})$ in 
Eq.~(\ref{theta2}) as $\dot \tau (\eta )$ in the integrand of
Eq.~(\ref{alpha_n_npr}) even though ${\bf x} = {\hat n} (\eta_0-\eta)$.
We have also ignored inhomogeneities in the free electron density 
along different directions on the sky since this will only give a 
higher order correction to FR.
The limits of the $\eta$, $\eta'$ integrations are from the
time that the photon last scattered to the present time. In general,
the last scattering time will be different for different photons but,
in the thin LSS approximation, the initial time will be taken to be
$\eta_*$, the epoch at which the ``visibility function'', 
$g(\eta) \equiv {\dot \tau} e^{-\tau}$, is maximum.

Statistical isotropy implies that 
$\langle \alpha ({\hat n}) \alpha ({\hat n}') \rangle$ must be a 
function of ${\hat n} \cdot {\hat n}'$. This can also be seen directly 
by writing Eq.~(\ref{alpha_n_npr}) as
\be
\langle \alpha ({\hat n}) \alpha ({\hat n}') \rangle = 
\left(\frac{3\lambda_0^2}{16 \pi^2 e} \right)^2
\int \frac{k^2 dk}{(2\pi )^3} S(k) \int d\eta \int d\eta'
 {\dot \tau}(\eta) {\dot \tau}(\eta') 
\left[{\hat n}\cdot {\hat n}' - \partial_{k \eta} \partial_{k \eta'} \right]
\int d^2 {\hat k} e^{-i{\bf k}\cdot {\hat n}\eta} e^{+i{\bf k}\cdot {\hat n}'\eta'} \ ,
\label{alpha_n_npr_1}
\ee
and using
\be
\int d^2 {\hat k} e^{-i{\bf k}\cdot {\hat n}\eta} 
    e^{+i{\bf k}\cdot {\hat n}'\eta'} =
4\pi \sum_l (2L+1) j_L(k\eta) j_L(k\eta') P_L(\hat n \cdot \hat n') \ .
\label{fourier-bessel}
\ee
where $j_L(x)$ are Bessel functions and $P_L(x)$ are Legendre
polynomials.
Hence, $\langle \alpha ({\hat n}) \alpha ({\hat n}') \rangle$  
depends only on ${\hat n} \cdot {\hat n}'$ as expected. As a consequence,
the correlator of $\alpha$'s can be expanded into Legendre series
\be
\langle \alpha ({\hat n}) \alpha ({\hat n}') \rangle = 
\sum_L {(2L+1)\over 4\pi} C_L^{\alpha \alpha} 
P_L(\hat n \cdot \hat n')
\label{legendre}
\ee
and the FR correlators can also be written as
$\langle \alpha^*_{LM}  \alpha_{L'M'}  \rangle = 
C_L^{\alpha \alpha} \delta_{LL'} \delta_{MM'}$,
where $\alpha_{LM}$ are the coefficients in the spherical harmonic
decomposition of the FR angle,
\begin{equation}
\alpha ({\hat n}) = \sum_{L,M} \alpha_{LM} Y_{LM}({\hat n}) \ .
\end{equation}

Now combining (\ref{alpha_n_npr_1}) and (\ref{fourier-bessel}), and 
introducing $j_L'(x) \equiv 
\partial_x j_L(x)$, we can write
\ba
\nonumber
\langle \alpha ({\hat n}) \alpha ({\hat n}') \rangle &=& 
\left(\frac{3\lambda_0^2}{16 \pi^2 e} \right)^2 {2\over \pi }
\int k^2 dk S(k) \int d\eta \int d\eta'
 {\dot \tau}(\eta) {\dot \tau}(\eta') \\
 &\times& \sum_L  {(2L+1) \over 4\pi} 
 \left[ 
 (\hat n \cdot \hat n')  P_L(\hat n \cdot \hat n')  j_L(k\eta) j_L(k\eta') -  j_L'(k\eta) j_L'(k\eta') P_L(\hat n 
\cdot \hat n') 
 \right] \ .
\label{alpha_n_npr_2}
\ea
Applying the recursion relation 
\be
(L+1)P_{L+1}(x)=(2L+1)xP_L(x)-LP_{L-1}(x)
\ee
to the 
$(\hat n \cdot \hat n')  P_L(\hat n \cdot \hat n')$ term in 
(\ref{alpha_n_npr_2}) results in
\ba
\nonumber
\langle \alpha ({\hat n}) \alpha ({\hat n}') \rangle &=& 
\left(\frac{3\lambda_0^2}{16 \pi^2 e} \right)^2 {2\over \pi }
\int  k^2 dk S(k) \int d\eta \int d\eta'
 {\dot \tau}(\eta) {\dot \tau}(\eta') \sum_L  {(2L+1) \over 4\pi} P_L(\hat n \cdot \hat n') \\
 &\times& 
 \left[ 
{L \over 2L+1} j_{L-1}(k\eta) j_{L-1}(k\eta') + {L+1 \over 2L+1} j_{L+1}(k\eta) j_{L+1}(k\eta') - j_L'(k\eta) 
j_L'(k\eta') 
 \right] \ .
\label{alpha_n_npr_3}
\ea
In analogy with the way $C_L$'s are evaluated for CMB, it is convenient 
to introduce ``transfer'' 
functions ${\cal T}_L(k)$ and ${\cal T}_L^{(1)}(k)$ defined as
\ba
\nonumber
{\cal T}_L(k) &\equiv&  
\int_{\eta_*}^{\eta_0} d\eta  ~ {\dot \tau}(\eta) j_L(k(\eta_0-\eta)) \\
{\cal T}_L^{(1)}(k) &\equiv&  
\int_{\eta_*}^{\eta_0} d\eta  ~ {\dot \tau}(\eta) j_L'(k(\eta_0-\eta))
\ea
where, as defined above, $\eta_*$ is the epoch at which the visibility
function is maximum and $\eta_0$ is the
present epoch. Using these transfer functions in (\ref{alpha_n_npr_3}) 
and comparing to (\ref{legendre}) allows us to write 
\begin{eqnarray}
C_L^{\alpha \alpha} = {2\over \pi } \int {dk \over k} \Delta^2_M(k)
\left[ 
{L \over 2L+1} ({\cal T}_{L-1}(k))^2 + {L+1 \over 2L+1} ({\cal T}_{L+1}(k))^2 - ({\cal T}_L^{(1)}(k))^2
 \right] \ .
\label{c_ell} 
\end{eqnarray}
The function ${\cal T}_L^{(1)}$ can be expressed in terms of
${\cal T}$ functions by using the relation
\be
j_L'(x)=\frac{1}{2L+1}\left[Lj_{L-1}(x)-(L+1)j_{L+1}(x)\right] \ .
\ee
This gives
\be
{\cal T}_L^{(1)} = 
\frac{1}{2L+1}\left[L{\cal T}_{L-1}-(L+1){\cal T}_{L+1}\right] \ .
\ee
The functions ${\cal T}_L(k)$ are independent of the magnetic field, 
and can be easily evaluated numerically using the ionization history from 
CMBFAST. 

Eq.~(\ref{c_ell}) is our final result for the FR
correlation function. It will be useful in Sec.~\ref{sec:thin} where we 
find $C_l^{BB}$ in the thin LSS approximation. However, the result is 
not useful in the general case of a thick LSS because then $C_l^{BB}$ 
is not simply related to $C_L^{\alpha\alpha}$.

\section{Faraday Rotation of CMB in thin LSS approximation}
\label{sec:thin}

In the limit of instant last scattering one assumes that all of the 
polarization was generated at the peak of the visibility function. 
Since we are interested specifically in the FR effects, 
we will neglect primordial tensor modes and any actively sourced vector 
and tensor modes (including those sourced by magnetic fields) so that only 
$E$ mode is produced at the instant of last scattering. 
At subsequent times, because 
of the residual presence of charged particles, some of this $E$ mode 
will be Faraday rotated into $B$ mode. To estimate this effect, we can 
start with Eq.~(6) of \cite{Kamionkowski_08} (same as Eq.~(20) of 
\cite{2009PhRvD..80b3510G}) which gives the B-mode coefficients
\be
B_{lm}=2(-1)^m\sum_{LM}\sum_{l_2 m_2}\alpha_{LM} E_{l_2 m_2} 
\xi_{lml_2m_2}^{LM}H_{ll_2}^L \ ,
\label{eq:blm}
\ee
where $\xi_{lml_2m_2}^{LM}$ and $H_{ll_2}^L$ are defined in 
terms of Wigner $3-j$ symbols as \cite{2009PhRvD..80b3510G}
\ba
\xi_{lml_2m_2}^{LM} &\equiv& (-1)^m \sqrt{ (2l+1)(2L+1)(2l_2+1) \over 4\pi}
\left(
\begin{array}{ccc}
l  & L  & l_2  \\
-m  & M  & m_2    
\end{array}
\right) \\
H_{ll_2}^L &\equiv& 
\left(
\begin{array}{ccc}
l  & L  & l_2  \\
2  & 0  & -2    
\end{array}
\right) \ ,
\ea
and the summation is restricted to be only over even $L+l_2+l$.
From the above, we can derive the expression relating $C_l^{BB}$ to $C_l^{EE}$:
\ba 
\nonumber
\langle B^*_{l'm'} B_{lm} \rangle &=& 4 \sum_{LM} \sum_{L'M'} \sum_{l_2 m_2}\sum_{l_2' m_2'}
 \xi_{lml_2m_2}^{LM}H_{ll_2}^L  \xi_{l'm'l_2'm_2'}^{L'M'}H_{l'l_2'}^{L'}
 \langle \alpha^*_{LM} E^*_{l_2 m_2} \alpha_{L'M'} E_{l_2' m_2'} \rangle \\
&=& \delta_{ll'} \delta_{mm'} 4 \sum_L  {(2L+1)\over 4\pi} C^{\alpha \alpha}_L \sum_{l_2}  (2l_2+1) C^{EE}_{l_2} (H_{ll_2}^L)^2
\ea
which assumes a statistically isotropic stochastic magnetic fields with a FR angular spectrum 
$C_L^{\alpha \alpha}$ given by Eq.~(\ref{c_ell}). Therefore, the B-mode angular spectrum in the thin last scattering approximation is
\begin{equation}
C_l^{BB} = \frac{1}{\pi} \sum_L  (2L+1) C^{\alpha \alpha}_L 
    \sum_{l_2}  (2l_2+1) C^{EE}_{l_2} (H_{ll_2}^L)^2 \,.
\label{clbb}
\end{equation}

It is instructive to put the expressions for the CMB observables in a 
form that separates the well-established physics of FR 
of CMB polarization from the particular form of the magnetic field 
spectrum. For example, substituting (\ref{c_ell}) into (\ref{clbb}), 
we can re-write the latter as
\be
C_l^{BB} =  \frac{2}{\pi} \int {dk \over k} \Delta^2_M(k) W_l (k) \ ,
\label{clbb-thin}
\ee
where $W_l(k)$ are ``window functions'' defined as
\be
W_l(k) =
4 \sum_{l_1 L} {(2l_1+1)(2L+1) \over 4 \pi}  (H_{ll_1}^L)^2 C_{l_1}^{EE} 
 \left(
{L\over 2L+1} [{\cal T}_{L-1}(k)]^2 + {L+1 \over 2L+1}[{\cal T}_{L+1}(k)]^2 -[{\cal T}^{(1)}_L(k)]^2 
 \right) \ .
\label{window-thin}  
\ee
They describe the amount of power a given wavelength $k$ of the magnetic 
field spectrum contributes to a given angular scale $l$ of the B-mode 
polarization spectrum. We note that Eq.~(\ref{clbb-thin}) relating 
$C_l^{BB}$ to the magnetic spectrum is formally independent of the 
thin LSS approximation -- the approximation is used in the calculation 
of the window functions.
Transfer functions ${\cal T}_L(k)$ can be found numerically using 
the differential optical depth calculated in CMBFAST~\cite{cmbfast}. Having 
evaluated the window functions once, one can store them and use 
(\ref{clbb-thin}) to calculate $C_l^{BB}$ for different choices of 
the magnetic spectrum.

\section{Exact (thick LSS) treatment of Faraday Rotation}
\label{sec:exactthick}

The thin LSS approximation decouples the process of generation of the 
CMB polarization by Thomson scattering from its subsequent FR by 
magnetic fields. It effectively assumes that the background E-mode 
polarization on all angular scales was created at a single instant 
in time corresponding to the peak of the visibility function. While 
this may suffice for order of magnitude estimates, polarization in 
different parts of the sky was created at different times and any 
choice of a single time is essentially arbitrary. Furthermore, the 
amount of the FR strongly depends on the choice of the initial instant, 
since the amplitude of the rotation is directly proportional on the 
rapidly decreasing free electron density. In this Section we derive 
the exact FR window functions, denoted by ${\tilde W}_l(k)$, by solving 
the radiative 
transport equations for the generation and propagation of CMB polarization 
in the presence of FR by stochastic magnetic fields. The form of 
Eq.~(\ref{clbb-thin}) relating $C_l^{BB}$ to the magnetic spectrum will 
be the same.

The evolution of CMB Stokes parameters is described by Chandrasekhar's 
radiative transport equations~\cite{chandra}. In 
the absence of FR, equations for the ${\bf q}$ Fourier mode of linear 
polarization parameters $Q$ and $U$ are~\cite{HuWh97}
\be
{\dot P_\pm}+iq\mu P_\pm = -{\dot \tau}P_\pm + S_\pm    \ ,
\label{transport}
\ee
where $P_\pm({\bf q},{\hat n},\eta) = Q\pm iU$, ${\hat n}$ is the direction of the line of sight, $\mu={\hat q} \cdot {\hat n}$, and
\be
S_\pm = {\dot \tau} \sqrt{6} \sum_{a=-2}^2 P^{(a)}({\bf q},\eta) \ _{\pm2}Y_{2a}(\hat n) \sqrt{4\pi \over 5} \ .
\ee
In the above, $_{\pm2}Y_{lm}(\hat n)$ are spin-2 spherical harmonics, and $P^{(a)}({\bf q},\eta)=[\Theta_2^{(a)}-\sqrt{6} E_2^{(a)}]/10$, where $\Theta_2^{(a)}$ and $E_2^{(a)}$ are the quadrupole moments of the CMB temperature and E-mode polarization for scalar ($a=0$), vector ($a=\pm 1$) and tensor (a=$\pm 2$) modes. Assuming that polarization generated by vector and tensor sources is negligible, we have
\be
S_\pm = {\dot \tau} \sqrt{6} P^{(0)}({\bf q},\eta) \ _{\pm2}Y_{20}(\hat n) \sqrt{4\pi \over 5} \ .
\label{S-source}
\ee

FR rotates $Q$ into $U$, and $U$ into $Q$, leading to a 
new term on the right hand of (\ref{transport}) \cite{KosLo96}:
\be
{\dot P_\pm}+iq\mu P_\pm = -{\dot \tau}P_\pm  \mp 2i \omega_B P_{\pm} + 
S_\pm \ .
\label{odeforP}
\ee
where $\omega_B(\hat n,\eta)=f {\dot \tau} {\bf B}({\bf r})\cdot {\hat n}$, ${\bf r}=(\eta_0-\eta){\hat n}$, $f=3\lambda_0^2/(2\pi e)$, and $\lambda_0$ and ${\bf B}$ are the {\it comoving} wavelength and magnetic field strength. 
The ordinary differential equation (\ref{odeforP}) has the inhomogeneous
solution
\be
P_{\pm}=\int_0^{\eta_0} d\eta \  {\tilde s}_\pm({\bf q},\hat n,\eta) 
           e^{\mp 2i \int_{\eta}^{\eta_0} \omega_B d\eta'} \ ,
\label{general-P}
\ee
with
\begin{eqnarray}
{\tilde s}_\pm &=&  S_\pm e^{-\tau}e^{-iq\mu (\eta_0-\eta)} 
\nonumber \\
&=& -{\dot \tau} e^{-\tau} \sqrt{6} P^{(0)}({\bf q},\eta) \sum_l (-i)^l  
        \sqrt{4\pi(2l+1)} [\epsilon_l^{(0)}(q(\eta_0-\eta)) \pm 
          i \beta_l^{(0)}(q(\eta_0-\eta)) ]  \ _{\pm2}Y_{l0}(\hat n)  \ ,
\label{scalar-source}
\end{eqnarray}
and $\tau \equiv \int_{\eta}^{\eta_0} d\eta' \ {\dot \tau}$. In the above, we used the identity (Eq.~(16) of \cite{HuWh97})
\be
 -\sqrt{4\pi \over 5}  \ _{\pm2}Y_{20}(\hat n)  e^{i{\vec q} \cdot {\hat n}r} = \sum_l (-i)^l  
\sqrt{4\pi (2l+1)} [\epsilon_l^{(0)}(qr) \pm i \beta_l^{(0)}(qr)]  \ _{\pm2}Y_{l0}(\hat n)  \ ,
\ee
with (in what follows we will not need $\beta_l^{(0)}$)
\be
\epsilon_l^{(0)}(x) \equiv
\sqrt{{3\over 8}{(l+2)! \over (l-2)!}} {j_l(x) \over x^2} \ .
\ee
For small $\omega_B$, we can write (\ref{general-P}) as
\be
P_{\pm}({\bf q},{\hat n})=\int_0^{\eta_0} d\eta \  {\tilde s}_\pm({\bf q},\hat n,\eta) \left[1\mp 2i \int_{\eta}^{\eta_0} \omega_B d\eta' \right] \ .
\label{eq:los-small-w}
\ee
Next, we can use the total angular momentum formalism of \cite{HuWh97} to derive an expression for $C_l^{BB}$ in terms of the magnetic spectrum $S(k)$. From Eq.~(55) of \cite{HuWh97} we have
\be
P_{\pm}({\bf q},{\hat n})=\sum_l (-i)^l \sqrt{4\pi \over 2l+1}\sum_{m=-l}^{l} 
\left( E_l^{(m)}({\bf q}) \pm i B_l^{(m)}({\bf q}) \right) \ _{\pm2}Y_{lm}(\hat n) \ ,
\label{eq:tot-angular}
\ee
where  ${\hat q}={\hat z}$. Note that in Eq.~(55) of \cite{HuWh97} the sum over $m$ runs only from $-2$ to $2$ because these are the only modes that can be sourced by scalar, vector and tensor fluctuations in the metric. However, the FR effect on the propagation of photons is not via perturbations of the metric tensor. Hence, to stay general, we keep the sum to be over all $m$ modes. Inverting (\ref{eq:tot-angular}) and using (\ref{eq:los-small-w}) we obtain
\ba
B_l^{(m)}({\bf q}) 
&=& {1 \over (-i)^l} \sqrt{2l+1 \over 4\pi} \int d {\hat n} \int_0^{\eta_0} d\eta   
{\dot \tau} e^{-\tau} \sqrt{6} P^{(0)}({\bf q},\eta) \nonumber \\
&\times& \sum_{l_1} (-i)^{l_1}  \sqrt{4\pi(2l_1+1)} [\epsilon_{l_1}^{(0)}
\left(_{+2}Y_{l_10}[_{+2}Y_{lm}]^* + _{-2}Y_{l_10} [_{-2}Y_{lm}]^* \right) \int_{\eta}^{\eta_0} \omega_B d\eta'  \ ,
\label{blmk}
\ea
where we have assumed that FR is the only source of B-mode. The angular spectrum $C_l^{BB}$ can be written in terms of $B_l^{(m)}({\bf q})$ as \cite{HuWh97}
\be
(2l+1)^2 C_l^{BB} = 4\pi \int {d^3{\bf q} \over (2\pi)^3} \sum_{m=-l}^l \langle B_l^{(m)*}({\bf q}) B_l^{(m)}({\bf q}) \rangle \ .
\label{clbb-general}
\ee
Introducing $g(\eta)={\dot \tau} \exp (-\tau)$
and $X^{m}_{ll_1} \equiv _{+2}Y_{l_10}[_{+2}Y_{lm}]^* + _{-2}Y_{l_10} [_{-2}Y_{lm}]^*$, and substituting (\ref{blmk}) into (\ref{clbb-general}) we obtain
\ba
\nonumber
(2l+1) C_l^{BB} &=&  4\pi \int {d^3{\bf q} \over (2\pi)^3} 6 \int_0^{\eta_0} d\eta \ g(\eta) \int_0^{\eta_0} d\eta' g(\eta')
\langle P^{(0)*}({\bf q},\eta)  P^{(0)}({\bf q},\eta') \rangle \nonumber \\
&\times& \sum_{m=-l}^l  \sum_{l_1} \sum_{l_2} (-i)^{l_1} i^{l_2}  \sqrt{(2l_1+1)(2l_2+1)} 
\epsilon_{l_1}^{(0)}(q(\eta_0-\eta))\epsilon_{l_2}^{(0)}(q(\eta_0-\eta')) \nonumber \\
&\times&
\int d {\hat n} \int d {\hat n}'  X_{ll_1}^{m*}(\hat n) X_{ll_2}^m(\hat n') 
 \langle \int_{\eta}^{\eta_0} d\eta' \int_{\eta'}^{\eta_0} d\eta'' \omega_B(\eta'',{\hat n})  \omega_B(\eta''',{\hat n}') \rangle \ .
\label{clbb-thick}
\ea
Eq.~(\ref{alpha_n_npr_2}) for the equal time two-point correlation of 
rotation measure is easily generalized to the unequal 
time FR correlation case above. Namely, we have
\be
 \langle \int_{\eta}^{\eta_0} d\eta' \int_{\eta'}^{\eta_0} d\eta'' \omega_B(\eta'',{\hat n})  \omega_B(\eta''',{\hat n}') \rangle 
 = {2\over \pi} \int {dk \over k} \Delta^2_M(k) \sum_{L} \left({2L+1 \over 4\pi} \right) P_{L}({\hat n} \cdot {\hat n}') {\cal U}_{L}(k,\eta,\eta') \ ,
\ee
where
\be
{\cal U}_{L}(k,\eta,\eta')={L \over 2L+1} {\cal T}_{L-1}(k,\eta){\cal T}_{L-1}(k,\eta')
+ {L+1 \over 2L+1} {\cal T}_{L+1}(k,\eta){\cal T}_{L+1}(k,\eta') - {\cal T}_L^{(1)}(k,\eta){\cal T}_L^{(1)}(k,\eta')
\label{calU}
\ee
and the transfer functions are the same as before, except for the range 
of the time integration:
\ba
\nonumber
{\cal T}_L(k,\eta) &\equiv&  \int_\eta^{\eta_0} d\eta'  
                        {\dot \tau}(\eta') j_L(k[\eta_0-\eta']) \\
{\cal T}_L^{(1)}(k,\eta) &\equiv&  \int_\eta^{\eta_0} d\eta'  
                        {\dot \tau}(\eta') j_L'(k[\eta_0-\eta']) \ .
\ea
The ${\cal U}_L$ in (\ref{calU}) can be written as a sum of terms with separated $\eta$ and $\eta'$ dependencies:
\be
{\cal U}_{L}(k,\eta,\eta')=\sum_{c=1}^3 u^{(c)}_L(k,\eta) u^{(c)}_L(k,\eta') \ ,
\ee
where
\be
u^{(1)}_L=\sqrt{L \over 2L+1} {\cal T}_{L-1} \ , \ \
u^{(2)}_L=\sqrt{L+1 \over 2L+1} {\cal T}_{L+1} \ , \ \
u^{(3)}_L= i{\cal T}_L^{(1)}
\ee
We can also relate $\langle P^{(0)*}({\bf q},\eta)  P^{(0)}({\bf q},\eta') \rangle$ to the primordial curvature power spectrum $\Delta^2(q) $ via
\be
\langle P^{(0)*}({\bf q},\eta)  P^{(0)}({\bf q},\eta') \rangle =  q^{-3}\Delta^2(q) P^{(0)*}(q,\eta)  P^{(0)}(q,\eta') \ .
\ee
Putting it all in (\ref{clbb-thick}) and integrating over the angular dependence of ${\bf q}$, we obtain
\be
(2l+1)C_l^{BB} = {2 \over \pi} \int {dk \over k} \Delta^2_M(k) 
\sum_{l_1,l_2,L} \sum_{mM} {\cal Z}^{LM*}_{l_10lm}{\cal Z}^{LM}_{l_20lm}  {2 \over \pi} \int {dq \over q} \Delta^2(q) 
\sum_{c=1}^3 d_{l_1L}^{(c)*}(q,k)d_{l_2L}^{(c)}(q,k) \ ,
\label{clbb-thick-1}
\ee
where we have defined
\be
{\cal Z}^{LM}_{l_10lm} = (-i)^{l_1} \sqrt{(2l_1+1)} \int d {\hat n} \ X_{ll_1}^{m} Y^*_{LM} \ ,
\ee
and
\be
 d_{l_1L}^{(c)}(q,k) = \int_0^{\eta_0} d\eta \ g(\eta) \sqrt{6} P^{(0)}(q,\eta)
\epsilon_{l_1}^{(0)}(q(\eta_0-\eta))u^{(c)}_L(k,\eta)\,.
\label{thick-transfer}
\ee
Using the expression for the integral of a product of three spin-weighted spherical harmonics
\be
\int d\hat{n} _{a}Y_{l_1 m_1}(\hat n) _{b}Y_{LM}(\hat n) _{c}Y_{l_2 m_2}(\hat n)
=\sqrt{\frac{(2l_1+1)(2L+1)(2l_2+1)}{4\pi}}
\left(
\begin{array}{ccc}
 l_1 & L  & l_2  \\
 m_1  &  M  & m_2    
\end{array}
\right) 
\left(
\begin{array}{ccc}
 l_1 & L  & l_2  \\
 -a  & -b  & -c    
\end{array}
\right) 
\ee
and the orthogonality property of Wigner $3$-$j$ symbols~\cite{Varshalovich+88} , 
we have
\be
\sum_{mM} {\cal Z}^{LM*}_{l_10lm}{\cal Z}^{LM}_{l_20lm} = 4\delta_{l_1l_2} {(2l+1)(2l_1+1)(2L+1) \over 4\pi} ({\cal H}_{ll_1}^L)^2  \ \ {\rm if} \ l+l_1+L= {\rm even, \ and} \ 0 \ {\rm otherwise} \ .
\ee
Substituting this into (\ref{clbb-thick-1}) we can write
\be
C_l^{BB} =  
\frac{2}{\pi} \int {dk \over k} \Delta^2_M(k) {\tilde W}_l (k) \ ,
\label{ClBBthick}
\ee
with
\be
{\tilde W}_l(k)=4 \sum_{l_1L}{(2l_1+1)(2L+1) \over 4\pi} ({\cal H}_{ll_1}^L)^2 S_{l_1L}(k) \ ,
\label{thickwindow}
\ee
where $l+l_1+L={\rm even}$, and
\be
S_{l_1L}(k)={2 \over \pi} \int {dq \over q} \Delta^2(q) 
\sum_{c=1}^3 |d_{l_1L}^{(c)}(q,k)|^2  \ .
\label{eq:Sl1Lk}
\ee

The thick LSS window function in Eq.~(\ref{thickwindow}) is very similar
to the thin LSS window function given by Eq.~(\ref{window-thin}). Noting that
\be
C_{l_1}^{EE} =  \frac{2}{\pi} \int \frac{dq}{q} \Delta^2 (q)  \left[ \int_0^{\eta_0} d\eta ~ g(\eta) \sqrt{6} 
              P^{(0)}(q,\eta) \epsilon_{l_1}^{(0)} (q(\eta_0-\eta)) \right]^2 \ ,
\ee
we can see that
\begin{equation}
\frac{2}{\pi} \int \frac{dq}{q} \Delta^2 (q) \sum_{c=1}^3 
|d_{l_1 L}^{(c)} (q,k)|^2
\end{equation}
of the thick case becomes
\begin{equation}
C_{l_1}^{EE} 
 \left(
{L\over 2L+1} [{\cal T}_{L-1}(k)]^2 + {L+1 \over 2L+1}[{\cal T}_{L+1}(k)]^2 -[{\cal T}^{(1)}_L(k)]^2 
 \right)
\end{equation}
of the thin case if the function
$d_{l_1 L}^{(c)} (q,k)$ defined in Eq.~(\ref{thick-transfer}) is 
``factorizable'', that is if
\begin{equation}
d_{l_1 L}^{(c)} \to u_L^{(c)} (k,\eta_*) 
              \int_0^{\eta_0} d\eta ~ g(\eta) \sqrt{6} 
              P^{(0)}(q,\eta) \epsilon_{l_1}^{(0)} (q(\eta_0-\eta)) \ .
\end{equation}
Thus the thick LSS case reduces to the thin LSS case if we disregard 
the convolution in Eq.~(\ref{thick-transfer}). 

The convolution in Eq.~(\ref{thick-transfer}) will in general not be factorizable, \ie the different $q$ modes of the source E-mode polarization are not all created at a single time before FR took place. Instead, functions $d_{l_1 L}^{(c)}(q,k)$ determine the relative amount by which a given $q$-mode, projecting into multipole $l_1$ of $C_{l_1}^{EE}$, is distorted by the $k$-mode of the magnetic field projecting onto multipole $L$ of the FR distortion spectrum.

To evaluate ${\tilde W}_l(k)$ numerically, we modified CMBFAST 
to calculate sources $S_{l_1L}(k)$ on a grid in $L$ and $k$. In 
order to accurately account for magnetic fields on scales up to a 
given $k_{\rm max}$ in Mpc$^{-1}$, one needs to evaluate the source 
up to $L_{\rm max} \sim (10^4 {\rm Mpc})  k_{\rm max}$. We included 
the $l_1$ modes up to $6000$, and confirmed that it is more than 
sufficient for all $k$ and $L$ because of the exponential suppression 
of the source E-modes by the Silk damping. Time required for the 
evaluation of sources needed for ${\tilde W}_l(k)$  was $\sim 4000$ CPU-hours,
where the sources 
were sampled for $450 k$ bins. Once the sources are calculated and 
stored, the sums over $l_1$ and $L$ in (\ref{thickwindow}) are quick 
to perform. 

Even though computing the exact window functions ${\tilde W}_l(k)$ 
takes a non-trivial amount of CPU time, it only needs to be done once 
for a given cosmological model. We have made our window functions, evaluated 
using the $\Lambda$CDM model with WMAP7 best fit parameters \cite{wmap5_cosmology}, publicly available at 
{\tt http://www.sfu.ca/$\sim$levon/faraday.html}, along with a Fortran code that calculates 
$C_l^{BB}$ for a given $\Delta^2_M(k)$.

\section{Results}
\label{results}

\begin{figure}
\includegraphics[height=0.63\textwidth]{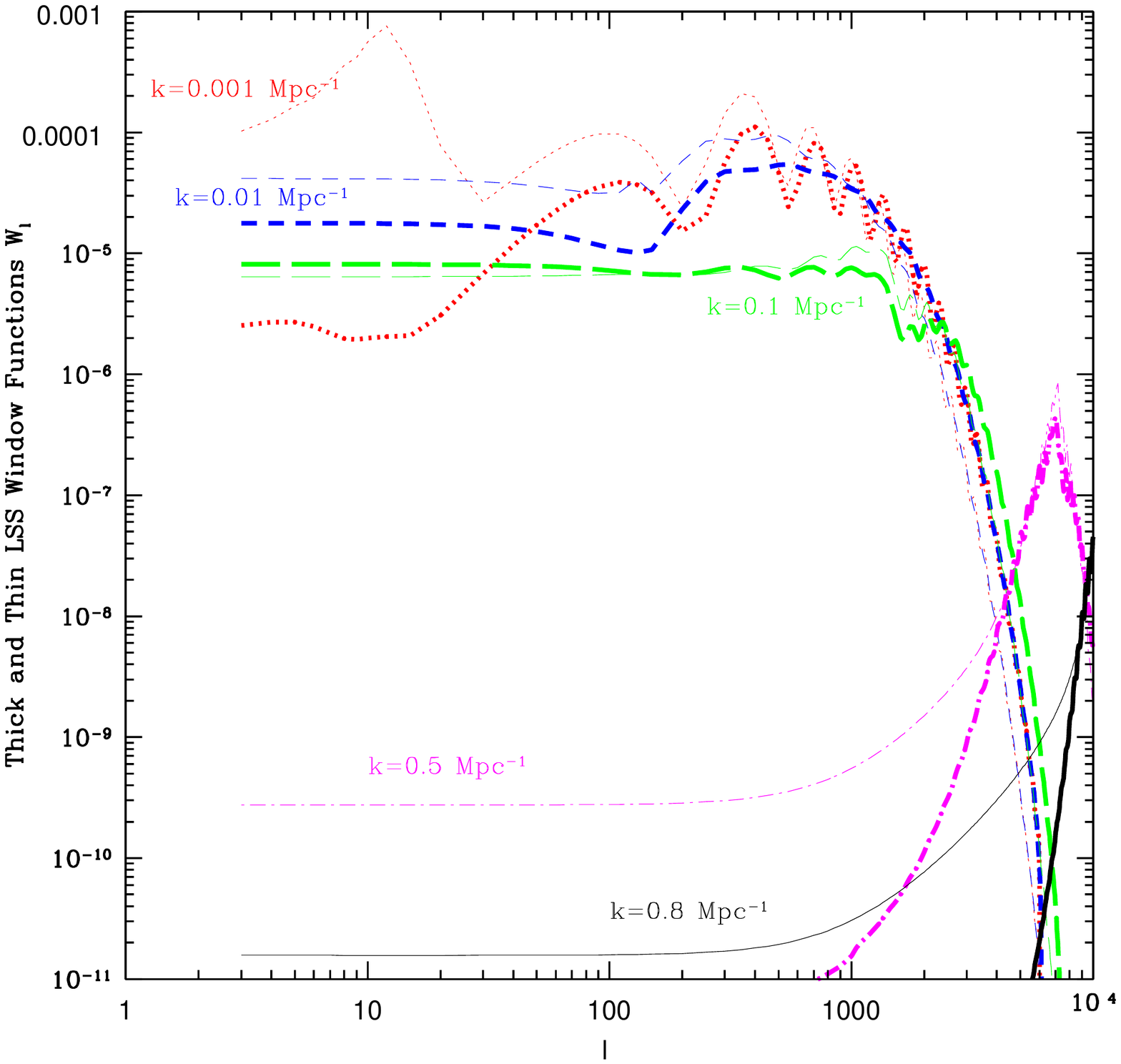}
\includegraphics[height=0.63\textwidth]{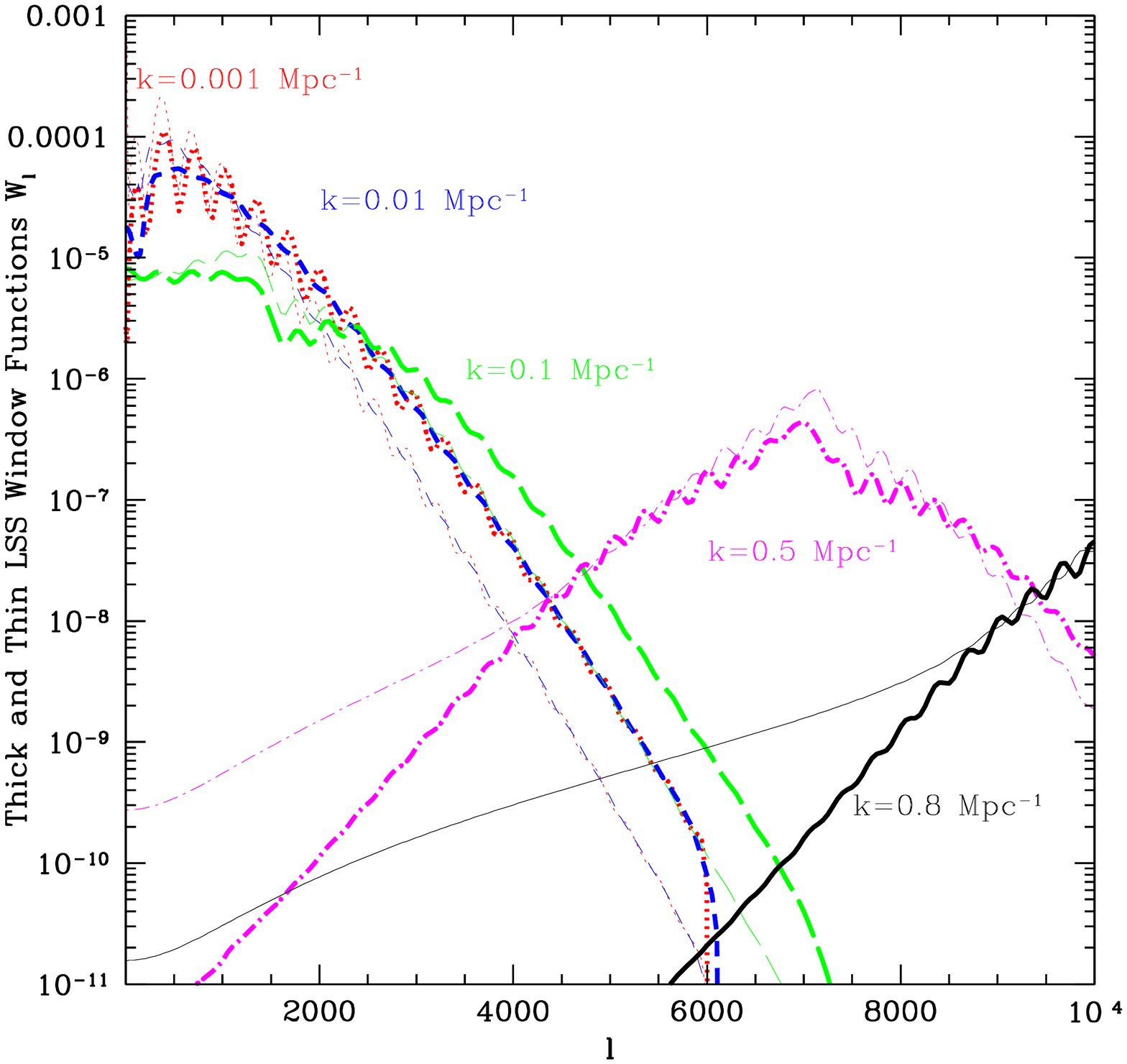}
\caption{Window functions $W_l(k)$ for different values of $k$ 
plotted vs $l$ as evaluated using the full LSS treatment (solid 
lines) and using the thin LSS approximation(dotted lines). These 
window functions prescribe the way in which a given Fourier mode 
$k$ of the stochastic magnetic field contributes to the multipole 
$l$ of $C_l^{BB}$. For example, $C_l^{BB}$ at $l \sim 8000$ is
sensitive to $\Delta_M^2$ at $k=0.5,~1~{\rm Mpc}^{-1}$ but not to
smaller $k$. The left and right panels show the same functions 
plotted on logarithmic and linear axis respectively.}
\label{fig:windows}
\end{figure}

In Fig.~\ref{fig:windows} we show window functions for several values 
of $k$ as a function of multipole $l$ for the thin (dotted lines) and thick 
(solid lines) LSS treatments discussed in previous sections. These 
results are independent of the spectral features of the magnetic field. 
Instead, the role of the window is to specify the extent to which a 
given Fourier mode $k$ of the stochastic magnetic field contributes to 
the multipole $l$ of $C_l^{BB}$. One can see that each window has a 
peak at $l$ approximately given by  $l = 10^4 k$, where $10^4$ is 
roughly the distance to LSS in Mpc. One can also see the oscillations 
which come from the acoustic oscillations in the E-mode spectrum.

Comparing the exact (thick LSS) windows with the ones obtained in the 
thin LSS approximation, we note that they have comparable shapes and 
amplitudes near their peaks, but differ significantly at $l$ away from 
the peaks. This difference comes because of the assumption made in the 
thin LSS approximation that all of the E-mode was produced at the 
same time, so that the FR occurs at the same rate on all scales. In 
reality, E-mode is rotated as it is being produced and, since the rate 
of FR depends on the rapidly decreasing free electron density, E-mode 
scales produced at different times are rotated at different rates.
 
\begin{figure}
\includegraphics[height=0.7\textwidth]{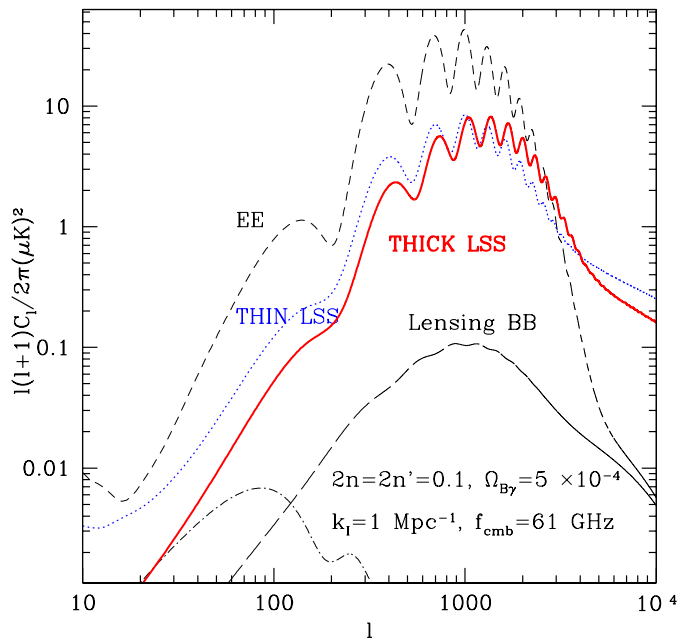}
\caption{
The CMB B-mode spectrum from Faraday rotation evaluated in the case of a nearly 
scale-invariant magnetic spectrum with $2n=2n'=0.1$, $k_I=1$ Mpc$^{-1}$, 
and $f_{\rm cmb}=61$~GHz, using the thin (blue dot) and exact 
(solid red) treatment of LSS. The value of the magnetic field energy fraction
$\Omega_{B\gamma}=5\times 10^{-4}$ corresponds to $B_{\rm eff} \approx 0.73\times 10^{-7}$ Gauss.
The black short-dash line is the input E-mode spectrum, the black dash-dot line is the contribution from
 inflationary gravitational waves with $r=0.1$, while the
black long-dash line is the expected contribution from gravitational lensing by large scale structure. 
}
  \label{fig:cl0}
\end{figure} 

Let us now focus on the B-mode spectra for specific choices of the 
magnetic spectrum. In Fig.~\ref{fig:cl0} we consider a nearly scale 
invariant magnetic spectrum with $2n=0.1$. The black dash line shows 
the input E-mode, while the blue dot and the red solid lines show 
the B-mode spectra obtained using the thin and exact 
treatment of LSS. One can 
see that the exactly calculated spectrum favors the power near the 
peak at the cost of the power around it. Note that in this case the 
shape of the B-mode is essentially a copy of the E-mode spectrum, 
except for the lack of exponential damping on small scales. While the EE correlations are suppressed
by the Silk damping, there is no exponential 
suppression of the FR generated small scale B-mode spectrum because the
magnetic field is correlated on small scales. For $2n=0.1$, both 
thin LSS and exact spectra have the asymptotic form of $l^{-1}$ at 
high $l$. More generally, the asymptotic exponent is $2n-1$, which
can be relatively large for stochastic fields {\it e.g.} for $n=5/2$,
$l^2 C_l^{BB} \propto l^4$ at large $l$.

The Silk damping dissipates the perturbations in the CMB temperature 
and E-mode on scales smaller than $9$ Mpc or so. In our formalism, 
it comes through the exponential suppression of the source function 
$P^{(0)}(k,\eta)$ in Eq.~(\ref{S-source}). However, the dissipation 
scale of (weak) magnetic fields at decoupling time is typically much 
smaller \cite{Jedamzik:1996wp,Jedamzik:2010cy} and for $k \lesssim 1$Mpc$^{-1}$ 
the magnetic field can be treated as a stiff source, {\it i.e.} we 
can safely assume that its evolution is independent of the perturbations 
in the photon-baryon fluid. Then, the only damping of the FR induced 
B-mode power is due to averaging over many random rotations along the 
line of sight. The functional form of this suppression is a power law 
and the exponent can be estimated by observing that a random superposition 
of $N$ perturbations along the line of sight leads to a statistical 
reduction in the amplitude of the observed anisotropy by a factor 
$1/\sqrt{N}$. For wavenumber $k$ we have $N \propto \Delta \eta k$, 
where $\Delta \eta$ is the period of time, comparable to the thickness 
of LSS, during which FR is efficient. Thus, the power spectrum on small 
scales, which is the square of the FR amplitude, is suppressed by $1/k$, 
which translates into the $1/l$ 
suppression of the angular spectrum.

\begin{figure}
\includegraphics[height=0.7\textwidth]{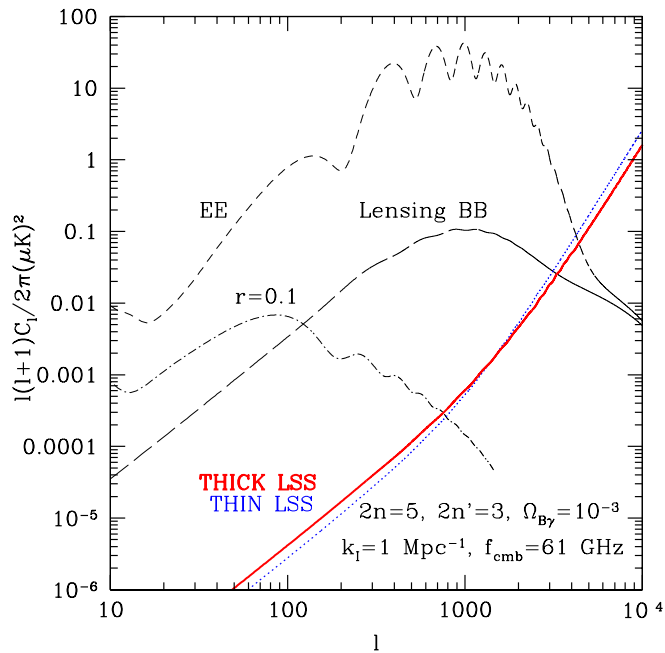}
\caption{
The CMB B-mode spectrum from Faraday rotation in the $2n=5$, $2n'=3$ case motivated by 
causally generated fields, with $k_I=1$ Mpc$^{-1}$ and $f_{\rm cmb}=61$~GHz.
The thin blue dot line shows the thin LSS calculation, while the exact (thick) LSS calculation 
is shown with solid red. The value of the magnetic field energy fraction, 
$\Omega_{B\gamma}=10^{-3}$, corresponds to $B_{\rm eff} \approx 10^{-7}$ Gauss. 
The other lines are the same as in Fig.~\ref{fig:cl0}.
}
  \label{fig:cl5}
\end{figure} 

\begin{figure}
\includegraphics[height=0.7\textwidth]{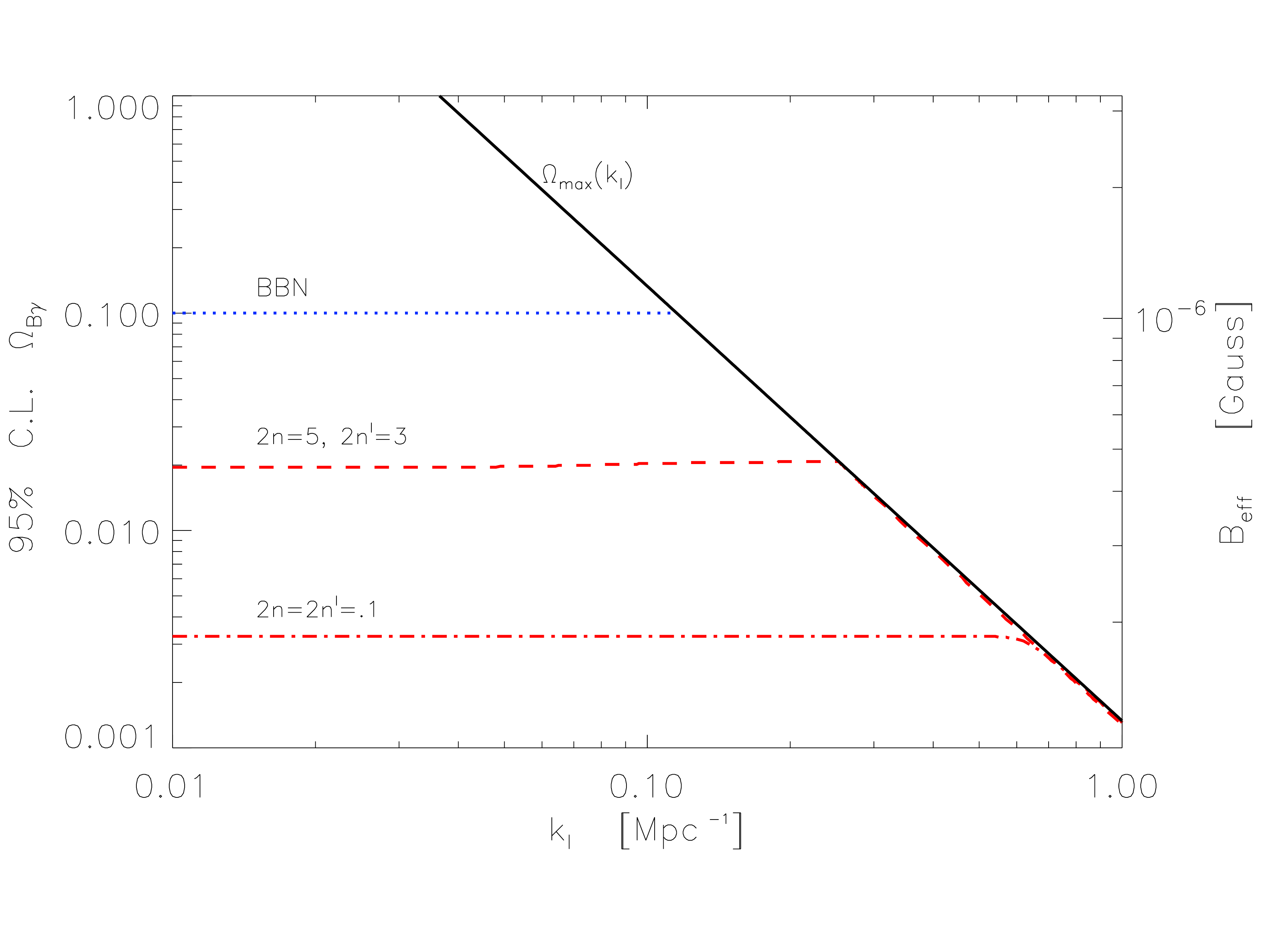}
\caption{Constraints from WMAP-7 year data on the magnetic field density 
$\Omega_{B\gamma}$, or analogously magnetic field effective amplitude 
$B_{\text{eff}}$ as defined by Eqs.~(\ref{def-beff}) and (\ref{beff-omega}), as a function of 
the inertial scale wavevector $k_I$. We consider two choices of magnetic 
spectrum, the case of a nearly scale-invariant spectrum $2n=2n'=0.1$ 
(dash-dot red line) and causal magnetic fields with 
$2n=5$, $2n'=3$ (dashed red line). The BBN 
constraints on the magnetic field density $\Omega_{B\gamma}<0.1$ 
is shown by the dotted blue line. 
At large $k_I$ the bound is set by the theoretical relation between $\Omega_{B\gamma}$ 
and $k_{\rm diss} \ge k_I$ given by Eq.~(\ref{kIBeff}) (black solid line).}
 \label{fig:constraints}
\end{figure} 

In Fig.~\ref{fig:cl5} we show the plot of the B-mode spectrum for 
the magnetic spectrum index $2n=5$ expected for causally generated 
magnetic fields~\cite{Jedamzik:2010cy}. In the figure we used 
$k_I=1$ Mpc$^{-1}$. For comparison, we show the B-mode from weak 
lensing and from gravity waves, as well as the E-mode spectrum. 
Note that at small angular scales (high $l$), the FR produced B-mode 
can dominate the signal as it keeps growing as $l^{2n-1}$.

At present, B-mode have not been detected; there are only upper bounds.
Still, even these weak bounds can produce constraints on the magnetic 
field fraction $\Omega_{B\gamma}$, and furthermore, the bounds will
improve rapidly as CMB observations are made on smaller angular scales.
We derive current constraints on the magnetic field energy density, 
$\Omega_{B\gamma}$, from the WMAP 7-year polarization data by
comparing magnetic field induced theoretical CMB B-mode power spectrum 
$C^{BB}_{\ell}$ as given by Eq.~(\ref{ClBBthick}), with the WMAP 
observed B-mode power spectrum using the $\chi^2$ statistics. We 
consider three WMAP frequency bands Q, V, and W corresponding to 
frequencies 41 GHz, 61 GHz, and 94 GHz respectively. The lower frequency 
bands, K and Ka, are foreground dominated and we do not include those in 
our analysis. We combine different frequency channels directly when evaluating 
the $\chi^2$ and, when evaluating the likelihood, we restrict the maximum value 
of $\Omega_{B\gamma}$ to the one given by Eq~(\ref{kIOmegaB}) for $k_I=k_{diss}$. 
We used WMAP data for $\ell>32$, above which the errors for 
individual $\ell$'s can be treated as uncorrelated. The maximum multipole considered 
in the analysis is $\ell=700$. For our analysis we 
considered two choices of theoretical magnetic spectral indices $(n,n')$, 
one corresponding to nearly scale-invariant spectrum $2n=2n'=0.1$, and 
the other corresponding to causal magnetic fields $2n=5$, $2n'=3$.  
Our analysis assumes that the magnetic field is the only source of the 
B-mode signal, and ignores the possibility of other sources of B-modes 
such as inflationary gravitational waves~\cite{2009AIPC.1141...10B}, weak 
gravitational lensing  of the CMB~\cite{2008arXiv0811.3916S}, and several 
other distortions of primary CMB along the line of sight 
(see Refs.~\cite{2010PhRvD..81f3512Y,2003PhRvD..67d3004H} for examples of 
such distortions).  

To obtain bounds on the magnetic field, we note that the CMB FR signature
constrains $\Delta_0^2$ defined in Eq.~(\ref{Delta20}), which is proportional
to $\Omega_{B\gamma}/\kappa$. Now $\kappa$ is defined in Eq.~(\ref{kappadef})
and depends on the dissipation scale, $k_{\rm diss}$, which can also be
related to $\Omega_{B\gamma}$ by using Eq.~(\ref{kIOmegaB}). Therefore for given
values of $n$ and $n'$,
$\kappa$ is a function of the inertial scale, $k_I$, and the energy
fraction in the magnetic field, $\Omega_{B\gamma}$. In Fig.~\ref{fig:constraints} we 
present the $95\%$ confidence level (C.L.). bounds on $\Omega_{B\gamma}$ for two 
choices of $(n,n')$: the nearly scale-invariant case, $2n=2n'=0.1$, and the causal case, 
$2n=5$, $2n'=3$. The bound in each case is 
presented as a function of $k_I$, which we treat as an unknown parameter in the model. 

The dependence of the constraint on $\Omega_{B\gamma}$ on $k_I$ in each
case can be readily understood. In the $2n=5$, $2n'=3$ case, $C_l^{BB}$ is independent of $k_I$ at very 
small $k_I$ because for positive $n$ the integral in Eq.~(\ref{ClBBthick}) 
is dominated by the contribution from $k \gg k_I$. The dependence of 
$C_l^{BB}$ on $k_I$ becomes stronger when $k_I$ gets closer to the range of 
scales with significant E-mode power ($k \sim 0.1 ~ {\rm Mpc}^{-1}$), which is encoded
in the shapes of the window functions. However, this happens to be just under the
maximum wavenumber constrained by WMAP's measurement of BB. Thus, 
WMAP is unable to probe the dependence on $k_I$, and the
constraint line in Fig.~\ref{fig:constraints} stays almost horizontal even at $k_I \sim 0.1$.
The condition $k_I \le k_{\rm diss} \propto \Omega_{B\gamma}^{-1/2}$ 
imposes its own upper bound on $\Omega_{B\gamma}$ at large $k_I$. 
Namely, at sufficiently large $k_I$, the constraint curve becomes independent of
the CMB constraints, and is a consequence of the constraint arising
from dissipation. The curve $k_I=k_{\rm diss}$ is also shown in Fig.~\ref{fig:constraints}.

In the scale-invariant limit, which is the $2n=2n'=0.1$ case in Fig.~\ref{fig:constraints}, 
the CMB constraint is independent of $k_I$ because all $k$ dependence effectively disappears. 
At large $k_I$, the bound is eventually dominated by the $k_I=k_{\rm diss}$
curve, which is independent of the CMB data.

We note that a CMB experiment which can measure B-modes at $\ell \sim 1000$ would be 
sensitive to changes in $k_I$ in the case of a causal spectrum. We will present forecasted
bounds from future CMB data in an upcoming paper \cite{amit-et-al}.

\section{Conclusions}
\label{conclusions}
Primordial stochastic magnetic fields may be produced in the early universe
during baryogenesis. Characterizing such primordial magnetic field is 
extremely valuable for probing early universe physics. In this paper we 
have calculated the effect of primordial magnetic fields on the CMB 
polarization. This study is especially timely in view of upcoming 
and next generation of CMB observations that are focused on measuring 
the polarization of the CMB. 

FR of existing E-modes of the CMB can generate parity odd 
B-modes whose amplitude will depend on the observation frequency bands.
We generalized the signatures previously obtained assuming ``thin'' last 
scattering surface by solving the full CMB radiative transport equation 
with FR taken into account. Comparing our full treatment 
with the thin last scattering approximation, we find the 
respective B-mode spectra have similar shapes but their magnitudes 
at a given multipole $\ell$ may differ by as much as a factor of a few 
depending on the primordial magnetic field spectrum and the angular
scale of observation. The well-established physics of FR of CMB can be 
encoded in window functions, $W_l(k)$,  which determine the relative amount 
that a given Fourier mode $k$ of the magnetic field contributes to the multipole 
$l$ of $C_l^{BB}$. The window functions are independent of the details of the 
magnetic field spectrum and only need to be computed once for a given 
cosmological model. We have evaluated them in the best fit $\Lambda$CDM 
model and made them available at {\tt http://www.sfu.ca/$\sim$levon/faraday.html}, 
along with a Fortran code that calculates $C_l^{BB}$ for a given magnetic spectrum.

FR can take a smooth E-mode field and distort it to create B-modes 
on arbitrarily small scales. For instance, on 
scales smaller than the Silk damping scale, the E-mode map is essentially 
homogeneous. However, FR creates E and B-mode inhomogeneities on sub-Silk 
scales by rotating different parts of the homogeneous E-mode patch in 
different random ways. As explained in the previous section, the power 
of these inhomogeneities is suppressed as $1/l$ due to random 
superposition of multiple small scale rotations along the same line 
of sight.

FR of the CMB due to magnetic fields has a distinct frequency 
dependence $(\sim\nu^2)$ which will allow it to be distinguished from other 
sources of CMB B-modes such as lensing, inflationary gravitational waves, 
topological defects \cite{Seljak:1997ii,Pogosian:2007gi}
and FR due to pseudo scalar 
fields \cite{Carroll:1998zi,Liu:2006uh,Pospelov:2008gg,Giovannini:2008zv}.
The growth $l^2 C_l^{BB} \propto l^{2n-1}$ at large $l$ is {\it also}
characteristic of a magnetic field and {\it will} help discriminate
primordial FR from other foreground contamination.
We have summarized current constraints on magnetic enegy $\Omega_{B\gamma}$ 
from WMAP7 based on the FR induced B-mode spectrum in 
Fig.~\ref{fig:constraints}. For comparison we have also shown the constraints 
on magnetic energy obtained by BBN. For a scale invariant magnetic field, 
the constraints from WMAP 7-year data are 2-orders of magnitude better than 
the BBN constraints. For a causal magnetic field with $(2n=5, 2n'=3)$, 
BBN constraints are weaker only by a factor of few. However,
note that the constraints from BBN are not expected to improve much in 
future, but future observations of CMB polarization at smaller angular
scales will significantly improve the constraints on magnetic field. We also
refer the reader to \cite{Paoletti:2010rx} for the most recent CMB bounds on
magnetic fields based signatures other than FR.

Although in this paper we have focused on B-mode power spectrum of 
CMB generated due to FR of the CMB polarization, 
recently it has been shown that the stochastic FR of 
polarization of the CMB couples different CMB angular modes, thus 
generating non-Gaussianity which can be seen in the CMB 
trispectrum~\cite{Kamionkowski_08,2009PhRvD..79l3009Y,2009PhRvD..80b3510G}. 
In a follow-up paper \cite{amit-et-al} we discuss the detectability of primordial magnetic 
field using such non-Gaussian features in the CMB.

\acknowledgments 
We are grateful to the WMAP collaboration, especially Eiichiro Komatsu and Michael Nolta, 
for providing the WMAP B-mode power spectra for individual frequency bands. 
We thank Karsten Jedamzik, Tina Kahniashvili, and other participants of
the Primordial Magnetism Workshop, 2011, at ASU for input, and Richard Battye and Yun Li 
for pointing out several inconsistencies in the earlier version of the paper.
LP is supported by a Discovery Grant from the Natural Sciences and Engineering 
Research Council of Canada, and acknowledges hospitality at Perimeter Institute 
for Theoretical Physics where part of this work was completed. 
A.P.S.Y. gratefully acknowledges funding support from NASA award number 
NNX08AG40G and NSF grant number AST-0807444.  TV is supported by the Department 
of Energy at ASU, and is grateful to the Institute for Advanced Study, 
Princeton (IAS) for hospitality. We also acknowledge the use of cluster 
computing at the IAS.

\bibliography{MagneticFields_references}

\end{document}